\input psfig.sty

\documentclass{aa}

\begin{document}

   \thesaurus{08.16.7, PSR\,J0537$-$6910, 03.09.1}
   
   \title{Search for the optical counterpart of the 16ms X-ray pulsar in the 
LMC
   \thanks{Based on observations collected at the European Southern 
    Observatory, La Silla, Chile}}

\author{R. P. Mignani\inst{1}, L. Pulone\inst{2}, G. Marconi\inst{2},
   G. Iannicola\inst{2} and P.A. Caraveo \inst{3,4}} 

   \offprints{R.P. Mignani (rmignani@eso.org)}

   \institute{
        ESA/ST-ECF, Karl-Schwarzschild-Str.2  D-85740 Garching, Germany -
         email: rmignani@eso.org
\and
Osservatorio Astronomico di Roma, Via di Frascati 33,   
I-00040 Monte Porzio Catone, Italy
\and         
Istituto di Fisica Cosmica ``G. Occhialini'', CNR, Via Bassini 15, 
I-20133-Milan, Italy 
\and 
Istituto Astronomico, Via Lancisi 29, I-00161-Rome, Italy 
}

 \date{Received 19 August 1999; accepted 23 November 1999 }

\titlerunning{ Optical obs. of PSR J0537$-$6910}
\maketitle

\begin{abstract} 

The 16ms X-ray  pulsar PSR\,J0537$-$6910 in  SNR N157B is the  fastest
known   isolated    (\underline{non-recycled}) pulsar   and,   with  a
rotational energy loss $\dot E  \sim 4.8 \times 10^{38}$ erg~s$^{-1}$,
is the  most energetic (together with  the Crab).   Here we report the
results of optical observations of the  field, recently performed with
the  SUSI2 camera of   the NTT.  Few objects   are observed inside the
$\simeq$ 3\arcsec~  X-ray   error circle  but none   of  them  can  be
convincingly associated to the  pulsar, which appears undetected  down
to $V \sim 23.4$.  With a corresponding optical luminosity $L_{\rm opt} 
\le 1.3 \times 10^{33}$ erg~s$^{-1}$, PSR\,J0537$-$6910 is, at best, comparable to the other very young pulsars Crab and
PSR\,B0540$-$69. 

   \end{abstract}

%

\section{Introduction}

PSR\,J0537$-$6910 is a young, fast,  X-ray pulsar, recently discovered
at the  center of  the LMC  supernova remnant N157B,   close to the 30
Doradus star forming  region.  N157B belongs   to the class of the  so
called Crab-like  supernova  remnants, or  plerions,  characterized by
non-thermal  spectra and   a centrally-filled radio/X-ray   morphology
probably due to  the presence of a  synchrotron nebula powered by  the
relativistic wind from a young, energetic, pulsar. Apart from the Crab
Nebula, so far  only three other plerions were  known to host pulsars,
namely: SNR0540$-$69 (PSR\,B0540$-$69), also  belonging to  30 Doradus
complex    and   located    15\arcmin~     from   N157B,    MSH$-$15-52
(PSR\,B1509$-$58) and G11.2$-$0.3 (PSR\,J1811$-$1926).  Thus, with the
detection  of PSR\,J0537$-$6910, N157B represents the  fifth case of a
plerion/pulsar association.   \\ Pulsed  X-ray  emission at 16  ms was
serendipitously discovered  during  a RXTE/PCA observation  towards 30
Doradus    (Marshall et al.   1998).    Soon after,  the pulsation  was
detected in archived    1993   ASCA/GIS data and    additional
confirmations     came  from  BeppoSAX   (Cusumano     et al.   1998).
PSR\,J0537$-$6910 takes  over   the   Crab  (33 ms)  as   the  fastest
``classical'' (i.e.  not spun up by matter  accretion from a companion
star) pulsar.  \\  The pulsar was  identified  in the ASCA/GIS  with a
X-ray source  detected  at  the  center of N157B   and  resolved  in a
point-like component  (the    pulsar  and,  probably,   its associated
synchrotron nebula) plus an elongated feature, the origin of which is
still uncertain.    In  both RXTE and  ASCA   data, the pulse  profile
appears characterized by a sharp ($\sim 1.7$ ms~ FWHM) symmetric peak,
which shows no obvious evolution during the time interval between ASCA
and RXTE observations (3.5 yrs).  The  period derivative of the pulsar
($\dot P  \simeq 5  \times  10^{-14}$ s~s$^{-1}$),  obtained  from the
comparison of multi-epoch timing  (Marshall et al.  1998;  Cusumano et
al.  1998), gives a spindown age of $ \approx 5\,000$ yrs), similar to
the  age  of the  remnant  estimated by  Wang   \& Gotthelf (1998a), a
magnetic field of  $\approx  10^{12}~G$,  typical  for a  pulsar  this
young, and a rotational  energy loss $\dot  E \sim 4.8 \times 10^{38}$
ergs~s$^{-1}$.  Substantially the same  results were obtained from the
timing analysis of  ROSAT/HRI data (Wang \& Gotthelf 1998b).  \\
In radio, PSR\,J0537$-$6910 has been observed  between June and August
1998 using   the 64m  radio telescope  in  Parkes but it  has  not been
detected  down  to  an  upper limit  of   $F_{\rm 14.GHz} \sim  0.04$  mJy
(Crawford et al.  1998).  Although  not really compelling, the present
upper  limit suggests that  PSR\,J0537$-$6910 is  weaker in radio than
both the Crab pulsar and PSR\,B0540$-$69. 

\section{Optical observations}

While in radio PSR\,J0537$-$6910 is an  elusive target, in the optical
domain the  situation appears more promising.  \\  Up to now, three of
the five  pulsars younger than  $   10\,000$ yrs have  been  certainly
identified  in the optical (Mignani  1998), where they channel through
magnetospheric emission  $\sim  10^{-5}-10^{-6}$ of   their rotational
energy output.   Since  PSR\,J0537$-$6910 is   very  young ($   \simeq
5\,000$ yrs) and,  with the Crab, it has  the highest $\dot E$, it  is
natural to assume that  also in this case a  significant amount of the
rotating power     be radiated in the    optical.   However, given the
uncertain  dependance  of  the  optical  luminosity  vs.   the  pulsar
parameters, it  is   difficult to  make  a prediction   on the  actual
magnitude of PSR\,J0537$-$6910.   A possible estimate  can be obtained
by a straight  scaling of the Pacini's relation  (see e.g.   Pacini \&
Salvati 1987) i.e.  neglecting the dependance of the pulsar luminosity
on its unknown optical duty cycle.   This would yield $V \simeq 24.6$,
after  correcting for  the interstellar  absorption  $A_{\rm V} \sim 1.3$,
estimated  applying the relation of  Fitzpatrick (1986) with an
$N_{\rm H}
\sim 10^{22}$ cm$^{-2}$, measured by the X-ray spectral fittings (Wang
\& Gotthelf 1998a).    However, we note  that the  other  young ($\sim
2\,000$  yrs) LMC pulsar,  PSR\,B0540$-$69,  with a  factor 3  smaller
$\dot  E$, has  a  magnitude $V = 22.4$ with an $A_{\rm V} \sim 0.6$
(Caraveo et al. 1992).   \\
Although  PSR\,J0537$-$6910 is   still  undetectable   in radio,   its
detection  in   ROSAT/HRI data   (Wang   \&  Gotthelf  1998b)  reduces
significantly  its position uncertainty down   to  $\pm$ 3\arcsec  and
prompts the  search for its optical  counterpart.  The scientific case
appears  similar to the one of  PSR\,B0540$-$69,  also discovered as a
pulsating X-ray   source (Seward et al.    1984), also embedded   in a
supernova  remnant (SNR0540$-$69) and   tentatively identified  in the
optical without the aid of a  reference radio position (Caraveo et al.
1992).  \\ In the following, we describe the results of the first deep
imaging of the field of PSR\,J0537$-$6910, performed with the ESO/NTT. 

\begin{figure}
\centerline{\hbox{\psfig{figure=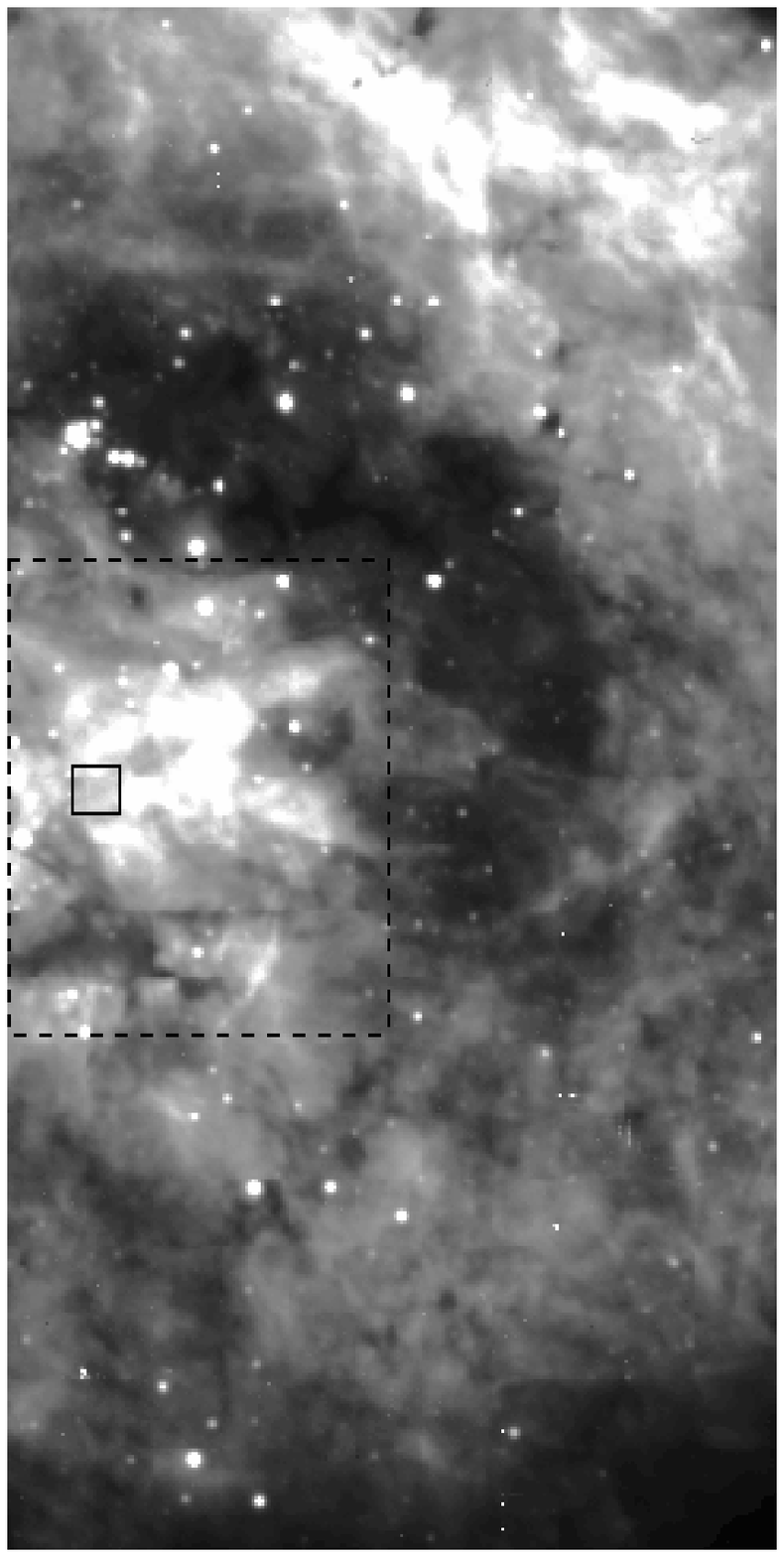,width=9cm,clip=}}}
\vspace{1.0cm}
\caption {SUSI2 $H_{\alpha}$ image of the 30 Doradus star forming region. The
location of  PSR\,J0537$-$6910 is near the center of the small 
$\simeq 20\arcsec \times 20\arcsec$  solid square. 
The larger ($0\farcm5 \times 0\farcm5$) dashed box marks the area in
which the search of stellar objects has been performed (see Sect. 2.2).}
\end{figure} 


\subsection{The data set}

\begin{figure}
\centerline{\hbox{\psfig{figure=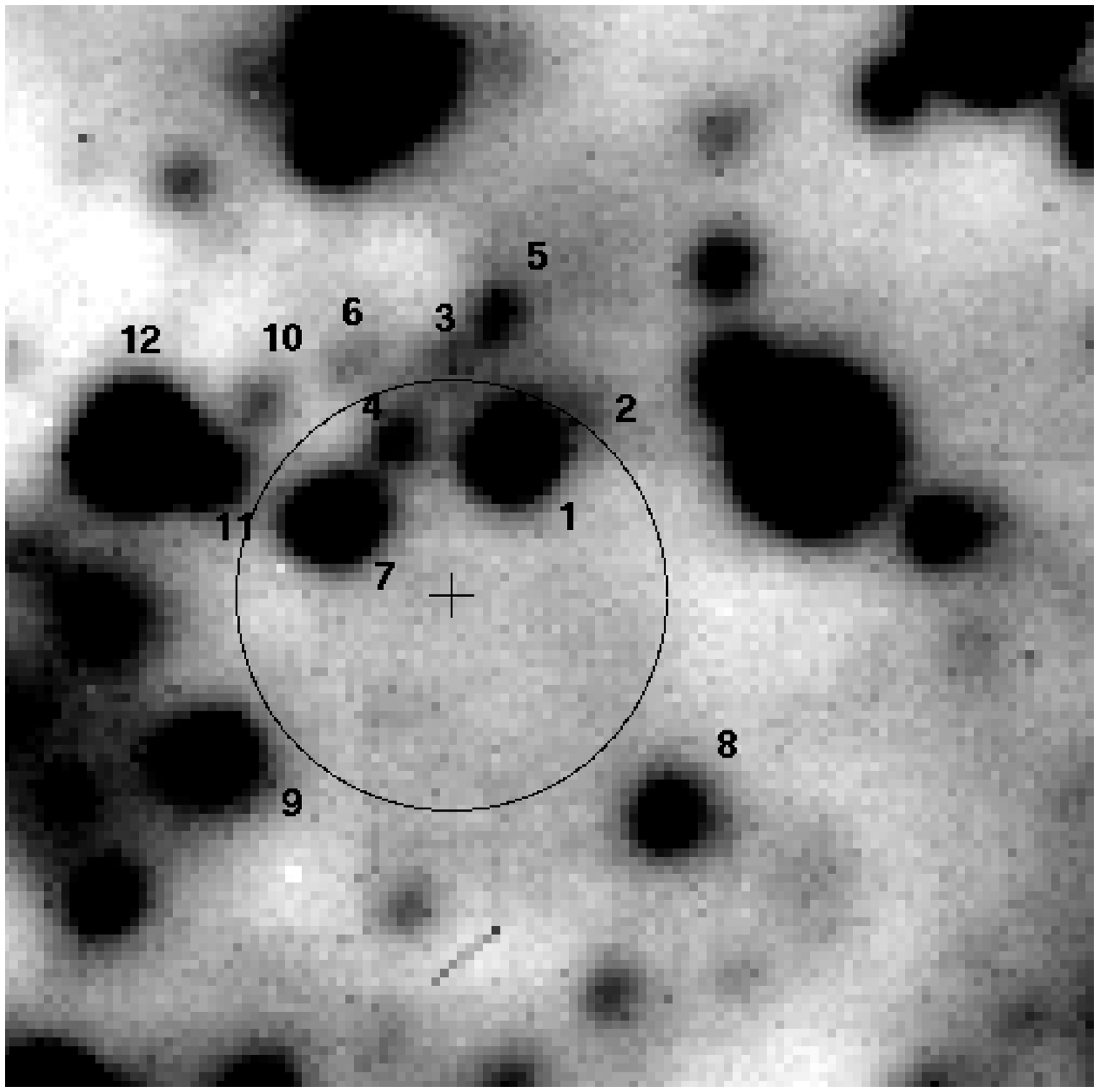,width=8cm,clip=}}}
\vspace{1.0cm}
\caption{ $\simeq 20\arcsec \times 20\arcsec$ I-band image of the 
PSR\,J0537$-$6910 surroundigs (North to the  top,  East to the   left). The
circle ($\simeq 4\arcsec$) marks the uncertainty region 
associated with the pulsar position as resulting from the
combination of the error ($\simeq
3\arcsec$) on the boresight-corrected ROSAT/HRI coordinates 
(Wang \& Gotthelf 1998b) and the global error budget of our astrometry. Objects detected
close/inside the error circle are labelled 1--12.  } 
\end{figure} 

The  field of PSR\,J0537$-$6910 has  been  observed in three different
runs between September and  November  1998 from the European   Southern
Observatory  (La Silla).    The observations  have  been performed  in
visitor mode with the NTT, equipped  with the second generation of the
SUperb Seeing Imager camera (SUSI2). The camera is a  CCD with a field
of view of $5\farcm5 \times  5\farcm5$, split in  two chips, and a
projected $2 \times 2$  binned pixel size of  0\farcs16.  The  two CCD
chips are physically  separated by a gap  $\simeq$ 100 pixels in size,
corresponding to  an effective sky masking of  $\simeq$  8\arcsec.  \\
Images were obtained in different wide-band filters ($B, V, I$) and in
the narrow-band $H_{\alpha}$, with  the available data set  summarized
in Table 1. 

\begin{table}[b]
\begin{center}
\caption{Summary of the available NTT observations of the PSR\,J0537$-$6910
field. Columns list the observing epochs, the
imaging device, the filters, the total exposure times in seconds 
and the average seeing conditions during each observation. 
}
\label{}
\begin{tabular}{|l|l|c|c|c|} \hline
{\em Date} & {\em Detector} & {\em Filter} & {\em T(s)} & {\em Seeing} \\
\hline
Sept 98 &  SUSI2 & V & 1800~s     & 1\farcs2    \\
        &  SUSI2 & I & 2400~s    & 1\farcs0    \\
        &  SUSI2 & $H_{\alpha}$ & 1200~s & 1\arcsec.0\\ 
Oct 98  &  SUSI2 & B & 3 $\times$ 500~s     & 1\farcs2 \\ 
Nov 98  &  SUSI2 & V & 2 $\times$ 900~s      &  0\farcs8\\ 
\hline
\end{tabular}

\end{center}
\end{table}

After the  basic    reduction steps    (bias subtraction,    flatfield
correction, etc.), single  exposures have  been combined and   cleaned
from cosmic ray hits by  frame-to-frame comparison.  The frames  taken
through  the same filter  have been  registered  with respect to  each
other  and combined  through   a median filter.  The  conversion  from
instrumental magnitudes to the  Johnson  standard system was  obtained
using a set  of primary  calibrators  from Landolt fields  observed at
different airmasses during each night.  The formal  errors in the zero
point of the calibration curves are $0.04$ magnitudes  in $V$ and $B$,
and $0.03$ in $I$.  \\ Astrometry on the field has been computed using
as a  reference   the coordinates and  positions  of  a set  of  stars
extracted from   the  USNO catalogue.    Then the  sky-to-pixel
coordinate transformation  has been  computed using the  ASTROM
software (Wallace 1990), yielding a final accuracy  of 0\farcs4 on the
astrometric  fit. \\  Fig.\,1 shows a   $1200 s$ exposure $H_{\alpha}$
image of  the 30 Doradus  region,  obtained through the  SUSI2 camera.
The solid square ($20\arcsec \times 20 \arcsec$), located close to the
maximum of the emission in the $H_{\alpha}$ band just at the center of
the   star    forming   region, includes   the     X-ray  position  of
PSR\,J0537$-$6910 (Wang \& Gotthelf  1998b).  A zoomed I-band image of
this area is shown in Fig.\,2 in negative greyscale.

\subsection{Results}

Few objects are  seen close or within the   X-ray error circle of  the
pulsar (Fig.\,2), including the moderately bright ($V \simeq 19$) star
\#1.  However, the crowding of the region, together with the irregular
sky background conditions,  prompted   us to apply   automatic  object
detection   routines to  search   for  additional, barely  detectable,
candidates.   

\begin{table}[h]
\begin{center}
\caption{Observed, not dereddened, magnitudes of the objects observed 
inside/close to the X-ray error circle. Objects are identified by
their labels in Fig.\,2. 
Columns list the $V$-band magnitudes, and colors. Attached photometric 
errors are $\simeq 0.02$ mag and $\simeq 0.1$ mag  at $V=18$ and
$V=21$, respectively.
}
\label{}

\begin{tabular}{|c|c|c|c|} \hline
{\em Id}  & {\em V} & {\em B-V} & {\em V-I} \\  
\hline
1&  18.8 & 0.38 & 0.45 \\
2&  21.2 & $>$1.9   & 0.85 \\
3&  20.4 & 0.40 & 0.02 \\
4&  20.7 & $>$2.4  & 0.72 \\
5&  19.9 & 0.32 & 0.11 \\ 
6&  20.5 & 0.58 & -0.70 \\ 
7&  20.0 & 0.54 & 1.58 \\ 
8&  19.8 & 0.72 & 0.62 \\
9&  19.8 & 0.95 & 1.11 \\
10& 20.8 & 0.21 & 0.74 \\
11& 20.0 & 0.35 & 0.36 \\
12& 17.4 & 0.13 & 0.27 \\ \hline
\end{tabular}

\end{center}
\end{table}

The object search in  the X-ray error circle  was thus performed using
the  ROMAFOT package  for photometry  in  crowded fields (Buonanno  \&
Iannicola 1989).  The ROMAFOT parameters were tuned to achieve in each
filter  a conservative $\ge 5\sigma$ object  detection above the local
background  level.  A   template PSF   was  obtained by   fitting  the
intensity  profiles of some  of  the brightest, unsaturated,  isolated
stars in the field with a Moffat function, plus a numerical map of the
residual  to better take into  account the contribution of the stellar
wings.   To allow   for an automatic  object   matching and  make  the
color-color analysis  faster, all the  images  have been  aligned to a
common reference  frame.  As a reference  for object detection we have
used our $I$-band image, where the effects of the local absorption are
reduced.  The master list of objects thus  created was then registered
on the  images taken in  $B$ and $V$ filters and  used as an input for
the fitting procedure.  A carefully check by eye has been performed in
order to  ensure that all  the stellar objects  found in the  $I$ band
were successfully fitted  in the other  images.   Apart from  the ones
labelled in  Fig.2,  no other  candidate  optical counterpart  to  the
pulsar has been clearly detected by our procedure.  We just report the
possible  presence  in the I-band  image  of an $\sim  22.3$ magnitude
object  (not recognizable   in Fig.2),   right   below the   detection
threshold  and  located nearly at the   center of the   error circle.
However,  the very low  significance of this detection  as well as the
lack  of color  information prevent us  to assess  the nature of  this
object and to speculate about a  possible association with the pulsar.
\\ The  properties of  objects  \#1-\#12,  i.e.  their  magnitudes and
colors  ($B-V$ and  $V-I$) are  summarized  in Table 2.   According to
their colors  and brightness, all  these objects are likely identified
as  young massive stars.   Fig.\,3 shows  the color--magnitude ($I$ vs
$V-I$)  diagram computed for a  sample of objects  selected in a $\sim
0\farcm5 \times 0\farcm5$  surrounding area (the  region marked by the
dashed box in Fig.\,1), together with the Zero Age Main Sequence track
estimated from  a suitable  chemical composition ($Z=0.008$, $Y=0.23$)
for the LMC  stellar population (Cassisi, private comm.).
The objects   labelled in Fig.\,2   and listed in  Table\,2  have been
marked  by   open diamonds.  Although    broadened by   the
interstellar   absorption and   shifted   redward by the  differential
reddening, the  color--magnitude diagram  (CMD) of  all  the stars  is
indeed consistent with  a young stellar population  main sequence.  
Thus, the  optical  counterpart of  PSR\,J0537$-$6910  is probably too
faint to  be  detected against   the high background  induced by   the
supernova remnant and the  absorption   of the embedding HII   region.\\
Using  the template $PSF$s computed from  each image, artificial stars
tests have been run to estimate the $V$, $B$, and $I$ magnitude limits
of our images.  With the flux  normalization left as a free parameter,
artificial stars have simulated  and added to the corresponding images
at  $100$   different positions randomly  selected   inside  the error
circle.  Thus, the detection algorithm has been run  in a loop for the
above number of trials, with the  flux normalization adjusted to allow
for a $3 \sigma$ detection in  each filter.  Averaging over the number
of trials, we   have found $3  \sigma$ detection  limits corresponding
(within $0.2$ mag)  to $B \simeq 23.2$, $V  \simeq 23.4$ and $I \simeq
22.4$, which we have taken as an indication of the limiting magnitudes
achievable in each band. 

\begin{figure}[h]
\centerline{\hbox{\psfig{figure=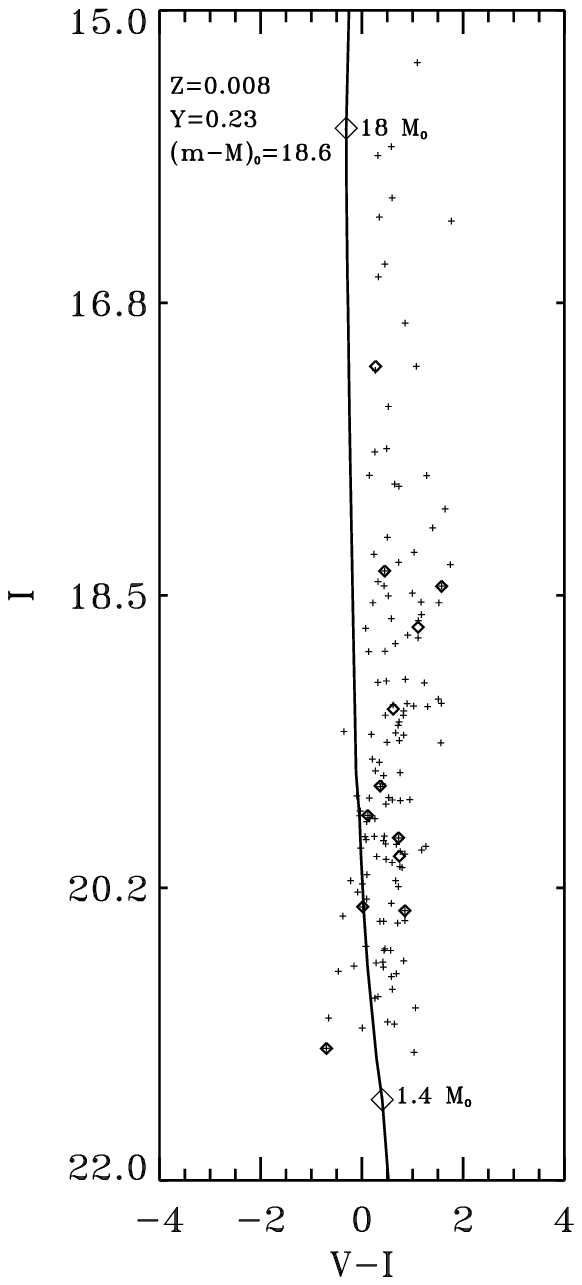,width=9cm,clip=}}}
\caption {Color--magnitude diagram of the objects found inside the
dashed box in Fig.\,1 (crosses) and the objects labelled in Fig.\,2
(open diamonds). The solid line in the plot corresponds to 
the Zero Age Main Sequence computed for stars with a chemical composition 
$Z\,=\,0.008, Y\,=\,0.23$ and masses in the range from $m=1.4 M_{\odot}$ to
$m=18 M_{\odot}$ (Cassisi, private comm.).}
\end{figure}

\section{Conclusions}

 We have performed deep optical observations to search for the optical
 counterpart of the isolated X-ray pulsar PSR\,J0537$-$6910.  However,
 none of  the objects detected close to/inside  the X-ray error circle
 stands out as a convincing candidate.  The marginal detection of a $I
 \sim 22.3$ object at the center of  the error circle must be regarded
 as  tentative  and is  in  need of  future confirmation.  The optical
 counterpart of PSR\,J0537$-$6910 is   thus unidentified down to a  $3
 \sigma$ limiting magnitude  of $\simeq 23.4$ in  V.  Our result is in
 agreement  with the upper   limits recently  derived by  Gouiffes \&
 \"Ogelman (1999) on the  pulsed optical flux.  \\  At the distance of
 47 kpc estimated for the host remnant N157B (Gould  1995) and for the
 assumed interstellar absorption ($A_{\rm V} \simeq 1.3$), our upper limit
 corresponds to  an  optical luminosity  $L_{\rm opt}   \le ~  1.3  \times
 10^{33}$  erg~s$^{-1}$.  This implies  that PSR\,J0537$-$6910  is, at
 best,   of  luminosity  comparable to    the ones   of   the Crab and
 PSR\,B0540$-$69, in  line with the  predictions  of Pacini's law.
 Although interesting, this upper limit is not stringent enough to put
 strong constraints on  the evolution of non-thermal  optical emission
 of young pulsars.  Together with the  recent upper limit obtained for
 PSR\,B1706$-$44 (Mignani et al. 1999), the measurement
 of the optical  luminosity of PSR\,J0537$-$6910  would be  crucial to
 smoothly  join the class of the  very young ($  \simeq 1\,000$ years)
 and bright   objects with the  class  of older,  Vela-like  ($ \simeq
 10\,000$  years) ones, for  which  the optical  output is $\simeq$  4
 orders of  magnitude  lower (Mignani  1998).  \\ As  in the   case of
 PSR\,B0540$-$69 (Shearer et al. 1994), time-resolved high resolution,
 imaging, possibly   exploiting  the  more   accurate  X-ray  position
 available from future  Chandra observations, would  be the best way
 to    pinpoint  and     identify     the  optical  counterpart     of
 PSR\,J0537$-$6910.

\begin{acknowledgements}
We acknowledge  the support software  provided by the  Starlink Project
which is funded by the UK SERC.  Part of the SUSI2 observations were
performed in guaranteed time as part of the  agreement between ESO and
the Astronomical Observatory of Rome.  Last, we would like to thank
the anonymous referee for his/her useful comments to the manuscript. 
  
\end{acknowledgements}


\begin{thebibliography}{References}
\bibitem[]{} Buonanno R.,  Iannicola G., 1989, PASP 101, 294
\bibitem[]{} Caraveo P.A., Bignami G.F., Mereghetti S., Mombelli
M., 1992, ApJ 395, L103
\bibitem{} Crawford F., Kaspi V.M., Manchester R.N., et al., 1998, 
MmSAI 69, 951
\bibitem[]{} Cusumano G., Maccarone M.C., Mineo T., et al., 1998, A\&A 333,  L55
\bibitem[]{} Fitzpatrick E.L., 1986, AJ 92, 1068
\bibitem[]{} Gouiffes K. \& \"Ogelman H., 1999, Proc. of IAU
Colloquium No. 177:  PULSAR ASTRONOMY - 2000 and Beyond, M.Kramer,
N.Wex and R.Wielebinski, eds. ASP Conference Series - in press
\bibitem[]{} Gould A., 1995, ApJ 452, 189
\bibitem[]{} Marshall F.E., Gotthelf E.V., Zhang W., et al., 1998,  ApJ 499, L179
\bibitem[]{} Mignani R.,  1998, Proc. of  Neutron Stars and Pulsars: Thirty 
years after the discovery.  N.~Shibazaki, N.~Kawai, S.~Shibata,  
T.~Kifune, eds. Universal Academic Press, Frontiers science series
n.24, p.335 
\bibitem[]{} Mignani R., Caraveo P.A., Bignami G.F., 1999,  A\&A
343, L5
\bibitem[]{} Pacini S., Salvati M., 1987,    ApJ    321, 447
\bibitem[]{} Seward F.D., Harnden F.R., Helfand D.J., 1984, ApJ
287, L19
\bibitem[]{} Shearer A., Redfen M., Pedersen H., et al., 1994, ApJ 423, L51
\bibitem[]{} Wallace P.T., 1990, Starlink User Note 5.11, Rutherford Appleton 
Laboratory
\bibitem[]{} Wang Q.D., Gotthelf E.V., 1998a,  ApJ 494, 623
\bibitem[]{} Wang Q.D., Gotthelf E.V., 1998b, ApJ 509, L109
\end{thebibliography}
\end{document}